\documentstyle[aps,preprint]{revtex}

\begin{document}
\draft
\title{ Simulating `Complex' Problems with Quantum Monte Carlo}
\author{
Lizeng Zhang$^{1,2}$, Geoff Canright$^{1,2}$ and Ted Barnes$^{1,3}$}
\address{
$^{1}$Department of Physics and Astronomy, \\
The University of Tennessee, Knoxville, Tennessee 37996-1200\\
$^{2}$ Solid State Division, Oak Ridge National Laboratory\\
Oak Ridge, Tennessee 37831\\
$^{3}$Physics Division, Oak Ridge National Laboratory\\
Oak Ridge, Tennessee 37831
}
\maketitle
\begin{abstract}
We present a new quantum Monte Carlo algorithm suitable
for generically complex problems, such as systems
coupled to external magnetic fields or anyons in two spatial
dimensions. We find that the choice of gauge plays a nontrivial
role, and can be used to reduce statistical noise in the
simulation. Furthermore, it is found that noise can be greatly
reduced by approximate cancellations between the phases of the
(gauge dependent) statistical flux and the external magnetic flux.
\end{abstract}

In quantum mechanics, although the Schr\"odinger equation
is complex, in many situations it can be studied in a real
representation. Such a representation can often greatly simplify
our analysis for those quantum systems. However, there are
some problems in which the complexity is unavoidable, in the sense
that such problems cannot be made real through a gauge transformation.
Examples of such generically complex problems are systems
with external magnetic fields or, in two dimensions, systems
of anyons.

Previous numerical investigations using quantum Monte Carlo (QMC)
methods have been mainly restricted to `real' systems.
In this paper, we introduce a QMC method appropriate for the
simulation of these generically `complex' systems.
Although our approach is quite general, we will illustrate our
method on a two dimensional square lattice system. In 2D, it is
well known that quantum particles may be anyons of arbitrary statistics,
which is defined by the phase $\theta$ of
the wavefunction obtained upon interchanging two identical
particles. Anyons can be represented as hard-core bosons
with appropriate (fictitious) attached flux tubes \cite{anyon}.
In this representation, a fermion is a boson with precisely one
flux quantum attached, whereas no flux is attached to a boson.
It is clear that either the statistical flux in the anyon
systems, or the magnetic flux when there is an external field,
gives rise to systems which are intrinsically complex.

For QMC simulations to be practical, it is important to achieve
a well controlled noise to signal ratio within a reasonable
computer simulation time. For this reason, the application of MC
techniques to fermionic systems at low temperature has been largely
infeasible due to the infamous `sign' problem
\cite{sign}. In complex problems, or in the complex representation
of the real problems that we will describe below, there is a
more general `phase problem' which can lead to equally
devastating effects on the simulations \cite{ZCB}.
However, our choice of two-dimensional problems allows us to explore a
continuum of possibilities between the well-controlled boson problem
and the relatively intractable problem of many fermions. Also, since it
is as easy to incorporate magnetic fields as fractional statistics,
this algorithm makes it possible to explore the interplay between
the two kinds of flux \cite{geoff}.

Our method is an unbiased form of the diffusion MC algorithm
\cite{Ted}. For illustration we consider a tight-binding
Hamiltonian of spinless particles on a 2D lattice
$H=H^{0} + H^{1}$ where
\begin{equation}
H^{0} = -t\sum_{\langle i,j\rangle }(e^{i\phi_{ij}}b^{+}_{i}b_{j} + H.c)
\;\;
\end{equation}
is the hopping term with $i,j$ representing site indices,
and $H^{1} = H^1(\{n_i\})$ contains interactions
as well as possible couplings to some external scalar potential.
In the above expression, we use a second quantized
notation in which $b^{+}_{j}$, $b_{j}$ are bosonic creation
and annihilation operators, subject to the hard-core
constraint so that the site occupation number operator
$n_{j}=b^{+}_{j}b_{j}$ has eigenvalue zero or one.
$\phi_{ij}$ is the phase associated with the hopping,
which may have contributions from the statistical flux as
well as from the external magnetic flux. The Schr\"odinger
equation of this system can be mapped onto a diffusion problem
by performing a Wick rotation $t \rightarrow -i\hbar\tau$,
resulting a diffusion equation
\begin{equation}\label{diffusion}
-\frac{\partial}{\partial \tau} |\psi(\tau)\rangle  = H |\psi(\tau)\rangle
\;\;,
\end{equation}
which can be simulated through an algorithm of weighted random
walks \cite{Ted}. Generalizations to the continuum or to particles
with spin are straightforward. Consider the wave function
$Q(Z,\tau) \equiv \langle Z|\psi(\tau)\rangle $,
where $\{ |Z\rangle  \}$, in which $H^{1}$ is diagonal,
spans the Hilbert space of
a system with a fixed number of particles $N_{p}$.
The discretized time evolution of $Q(Z,\tau)$ is given by
\begin{equation}\label{Qdisc}
Q(Z,\tau + h_{\tau})=W_{diag}(Z;\tau,\tau + h_{\tau} )\sum_{Z'}
h_{\tau}r_{Z' \rightarrow Z} W_{transition}(Z',Z;\tau,\tau+h_{\tau})
Q(Z',\tau) \;\; ,
\end{equation}
where the diagonal weight $W_{diag}$ and the off-diagonal
weight $W_{transition}$ are given by
\begin{equation}\label{Wd}
W_{diag}(Z;\tau,\tau+h_{\tau})=\exp\{-(V(Z,\tau)-
\sum_{Z'\neq Z}r_{Z \rightarrow Z'})h_{\tau} \}  \;\; ,
\end{equation}
\begin{eqnarray}
V(Z,\tau) = \langle Z|H^{1}|Z\rangle  \;\; , \;\;\;
r_{Z' \rightarrow Z} = \mid \langle Z|H^{0}|Z'\rangle  \mid \;\;,
\end{eqnarray}
and
\begin{equation}\label{Woffd}
W_{transition}(Z',Z;\tau,\tau+h_{\tau}) =e^{i\phi_{ij}} \;\; .
\end{equation}
One can verify that equation (\ref{Qdisc}) indeed reduces to the
diffusion equation in the limit of $h_{\tau} \rightarrow 0$.
The quantity $r_{Z' \rightarrow Z}$ can be interpreted as
the transition rate from state $|Z'\rangle $ to the state $|Z\rangle $
due to $H^{0}$, so that the probability $p_{Z' \rightarrow Z}$
for such a transition taking place during time $h_{\tau}$
is $h_{\tau} r_{Z' \rightarrow Z}$. Thus, starting with an
initial state, one can generate random walks according
to $p_{Z' \rightarrow Z}$, and compute the corresponding
weight factors.
The diagonal weights $W_{diag}$ contain two factors, representing the
external scalar potential $V$ and the kinetic energy. The off-diagonal
weights $W_{transition}$ are always a pure phase, with $\phi_{ij} =
\int_i^j \vec A \cdot d\vec{\ell}$ arising from a vector potential
$\vec A$ which represents the statistical and/or external flux.

We now consider the
calculation of physical observables at zero temperature,
in particular the ground state energy $E_{0}$. As
$\tau \rightarrow \infty$, one has
\begin{equation}\label{asympt}
|\psi(\tau)\rangle  \;\; = \;\; e^{-E_{0}\tau}|\psi_{0}\rangle
\langle \psi_{0}|\psi(\tau=0) \rangle
+ {\cal O}(e^{-(E_{1}-E_{0})\tau}) \;\; ,
\end{equation}
where $|\psi_{0}\rangle $ is the ground state of the system
and $E_{1}$ is the next lowest eigenvalue of $H$. The ground
state energy can be calculated from (\ref{asympt}) through, e.g.,
the logarithmic derivative of $\langle\psi(\tau)|\psi(\tau)\rangle$.
Actually, it is not necessary to compute
$|\psi(\tau)\rangle$ explicitly, since the ground state
energy can be determined directly from the weight
functions (\ref{Wd}) and (\ref{Woffd}). To see this,
note that the ground state energy $E_{0}$ may be extracted
equally well by multiplying to (\ref{asympt})
any state $|x\rangle$ not orthogonal to $|\psi_{0}\rangle$.
If we take $|x\rangle = \sum_{Z} |Z \rangle $ \cite{Ted},
we have
\begin{eqnarray}\label{Energy}
E_{0} = \lim_{\tau \rightarrow \infty} \frac{1}{\Delta \tau}
\ln\{\sum_{Z}Q(Z,\tau)/\sum_{Z}Q(Z,\tau + \Delta\tau) \}
\nonumber \\
= \lim_{\tau \rightarrow \infty} \frac{1}{\Delta \tau}
\ln\{\sum_{i=1}^{N_{rw}}W_{tot}^{i}(Z;0,\tau)
/\sum_{i=1}^{N_{rw}}W_{tot}^{i}(Z;0,\tau + \Delta\tau) \} \;\; .
\end{eqnarray}
where $N_{rw}$ is the total number of random walks
generated in the simulation, and the total weight function
$W_{tot} \equiv W_{diag}W_{transition}$.
In our actual simulations discussed below,
the random walks generated by $p_{Z' \rightarrow Z}$ are
sampled using the usual single move Metropolis method.
$E_{0}$ is computed through (\ref{Energy}) as
the mean of a large sample of random walks, while the
statistical fluctuations of such a sampling are
measured from its variance. To illustrate the algorithm and
for the purpose of investigating the effect of the `complexity', we have
not guided the random walks or done any other importance sampling.
For sufficiently small time step $h_{\tau}$, the systematic
error comes from the finite simulation time $\tau$ being
used during an actual simulation. The statistical noise,
as we show below, depends strongly on the
strength of the two sources of phase (the two fluxes) in the problem.
It also in general increases rapidly with
increasing $\tau$ as $\tau \rightarrow \infty$ \cite{ZCB,hetherington}.
Hence an accurate result for zero temperature often requires
a choice of finite $\tau$ which gives the best compromise
between the decreasing systematic error and the increasing
sampling error.  As an example, we show in Fig.1 the MC
result for three particles with various statistics on a
$4 \times 4$ lattice with free boundaries.
We see that the noise increases with
statistics approaching the fermion limit. For this system
there is a level crossing, near $\theta = 0.6$, at which the ground
state changes its symmetry. In this case,
one needs to multiply some additional symmetry breaking factors
to the weight $W^{i}_{tot}$ in (\ref{Energy}) to ensure
extracting the correct ground state properties; in the absence of
such a correction, our algorithm tracks the excited state of the
appropriate symmetry (as shown). The calculation illustrated
in Fig.1 is, to our knowledge, the first QMC simulation for
anyons.

It is clear that any physical observable computed should be
independent of the choice of the gauge. However, it is less
obvious that the statistical fluctuations associated with the
QMC simulations described above might not in fact be quite
sensitive to the gauge adopted. To illustrate the effect of
gauge choice on statistical fluctuations, let us consider
the `string gauge' \cite{claudius}. One may, for example, attach
a single string to each bare boson so that
there is a phase factor $e^{\pm i\theta}$  multiplying the
off-diagonal weight whenever a particle crosses a string
(or vice versa). (Since the strings have to end somewhere,
it is a nontrivial task to construct a gauge
theory for finite systems which are self-closed
(such as a torus). In the following we shall restrict ourselves
to problems with open boundary conditions where such
complications are not present.)
For bosons, all weights are positive, and no cancellation
occurs. As the statistical angle  $\theta$ is tuned away
from zero, $W_{tot}(Z;0,\tau)$ becomes complex. In general
we expect that the sum of the $W_{tot}^i$ in (\ref{Energy})
is also a complex number with a phase $\phi_{m}$.
One can study fluctuations by considering an additional
time increment $h_{\tau}$ after some (long) time
$\tau$. In this gauge the phase gained by a single MC
move can have $2N_{p}-1$ values, and has an angular variation
of amplitude $(2N_{p}-1)\theta$.
Here we see that phase fluctuations are minimized
in the small $\theta$ and/or small particle number limit.
For fermions ($\theta = \pi$), we have a situation where
the off-diagonal weights are real and fluctuate between
the positive and negative part of the
real axis, commonly known as the `sign problem'.

Now consider the gauge of $M$ strings per
particle. On a finite square lattice, the maximum number
of strings that can be attached to a particle is $4L$,
where $L$ is the linear size of the lattice. Each string
can share a statistical phase $\theta/M$. Depending on
how closely two particles are placed, the phase acquired
from a single hop of one particle
can be as large as $\theta/4$, or as small as $\theta/M$
(assume $M \geq 4$). Considering the dilute limit
(in which the lattice approaches a continuum) where
the probability for two particles coming close to each other
is negligible, the average phase fluctuation
caused by an additional MC move after time $\tau$
is of order $(2N_{p}-1)\theta/M$.
For $M=4L$, we see that for a given $\theta$,
the average phase fluctuation is reduced by a
factor of the order of the system size.

The simple argument given above should be treated with
some caution.  However, our argument does suggest,
in a intuitive way, that {\em fluctuations can be reduced
considerably by spatially
`spreading' the gauge}. (We note that there is some support
for this conclusion, in a different context, in Ref.~\cite{claudius}.)
Thus, from this point of view, it is quite plausible that
the optimum choice of the gauge which minimizes
the statistical fluctuations with a given number of
random walks would be the continuous gauge (CG)
[$A_\phi = (\theta/\pi r) \hat\phi$] which simply uses the
azimuthal increment $\delta\phi$ to determine the phase
increment associated with a hop. In this gauge the
change of phase associated with each MC move is minimum
on average. Furthermore, this argument suggests that the
CG will be most useful in the dilute limit, in which the angular
(and hence phase) increments can be made very small on average.
(Of course, in this limit, any method will find some relief
from statistical noise since the effect of statistics---e.g.,
the Fermi energy---is vanishing, regardless of gauge.)
We have carried out various simulations, using the diffusion
algorithm described above for free, spinless particles in 2D
to test the above conjecture.  In Fig.2 we show a comparison
of the single-string (1S) gauge and the CG,
for six spinless
fermions on a 20x20 lattice (filling 0.015). The CG estimate
is seen to be converging smoothly (although slowly, due to the
small excitation gap) to the exact result (calculated
independently) with increasing $\tau$, with bounded error estimates
for $\tau < 4.5$, while the 1S results give unbounded error in the same
range of $\tau$. We obtain this bound by examining the variance of the
distribution of (complex) weights to see if it
scales as $1/\sqrt{N_{rw}}$. This is essential since
even extremely noisy samplings can give---on occasion---a relatively
small sample variance.
Points which fail to meet this scaling criterion (and hence have a
sample variance which is not statistically reliable)
are marked with parentheses
in Figs.~2 and 3; the improvement obtainable by using the CG is
evident.

Now we explore the interplay between the statistical flux and
external flux. As discussed previously, the presence of external
flux may lead to large noise in QMC simulations, just as the
statistical flux can \cite{ZCB}. However, with the presence of
both fluxes, we have found that there is a nontrivial cancellation
between the two, resulting significant improvement in the signal
to noise ratio in QMC simulations.  Fig.3 shows results for the
simulation of 4 fermions  (treated in the CG)
on a $4\times 4$ lattice with an external magnetic field.
Without the field, the statistical noise is unbounded (by our
previous criterion) within the given computational time; as
the external field is increased (with the gauge choice such
that the statistical flux opposes to the external one), noise
is reduced progressively, such that, in the same simulation time,
the energy estimate converges to the exact result with much smaller
statistical error. Presumably the convergence rate is enhanced by
the larger gap in the case with flux; however, it is clear from the
figure that it is the statistical noise itself, arising from the large
phase fluctuations in the zero-field case, which renders the simulation
unmanageable. Finally, we note that increasing the field beyond the
optimal value again increases the statistical noise, so that it
resembles the simulation shown in Fig.3 for the zero-field case.

To summarize our results, we have introduced a new QMC
technique for simulations of generically complex problems.
Using a complex (anyonic) representation of the many
fermion system (which is not possible for QMC
in its real form), we find that statistical fluctuations
can be significantly reduced by an appropriate choice of
gauge. This method is presumably most effective in dilute
systems in which the average phase change in a single MC
move is small.  Furthermore, we find that there exist nontrivial
cancellations between statistical flux and external flux
for some external fields, resulting in considerable
reduction of the statistical error. This may be utilized in QMC
simulations for these systems.

There are some works in the literature exploring `complexity'
in QMC simulations \cite{complex}, most notably references
\cite{OCM} and \cite{qmcfq}. We would like to remark that
our algorithm is a general formalism which contains the
problems studied in these works as special
cases, and which allows one to simulate more general
systems, such as anyons (or fermions in the anyonic representation).
Furthermore, unlike the QMC methods commonly employed for lattice
problems which utilize
the Hubbard-Stratonovich transformation, this algorithm
requires no prior knowledge of single particle
states. This has thus the advantage that it permits
one to simulate problems for which such knowledge does
not exist or is difficult to compute (such as anyons).
Finally, we note that our method
requires small memory space on a computer so
that it is readily parallelized.

In closing, we wish to mention some physical problems
which might be investigated within this algorithm. First,
it is clear that our conclusions for fermions hold for anyons
in general.  The anyon problem
is of some physical interest both for its (possible)
superconductivity properties and for its relevance in the
fractional quantum Hall effect \cite{anyon}. This algorithm,
possibly enhanced by importance sampling in the generation of
random walks, should allow the study of anyon systems with sizes
not accessible to exact diagonalization.
We have also shown that representing fermions with a complex `anyon'
gauge can reduce the sampling noise. This reduction is however most
pronounced for dilute fermions, where other methods are already
successful. Hence we believe that a reformulation of the problem
in the continuum (rather than on a lattice) should be explored,
since this case the continuous gauge will have greatest
efficacy over the usual single-string gauge. Application
of our method to a continuum involves greater computational
demands but presents no conceptual difficulties.

Another problem involving external flux is the question of flux
quantization \cite{qmcfq,geoff2}, as measured by the dependence of
the ground-state energy upon a localized external flux.
We have already tested our method on the (hard-core) boson case,
on lattices up to 10$\times$10 in extent at 1/4-filling.
Interesting physical applications arise on extending the method
to bosons with disorder \cite{dirtybosons}, or to anyons \cite{geoff2};
in principle, fermions \cite{qmcfq} can also be studied with
this method as well. Finally, electrons in two dimensions in a
magnetic field is a problem of great current interest.
Our approach promises to be most useful at filling fractions
which give a commensuration of particles to flux, i.e., the
quantized Hall effect (QHE). It should
be straightforward to test for signs (such as a cusp in the energy) of
a QHE, for fermions or anyons on a continuum, or for particles on a
lattice \cite{nickread}.

We thank Steve Girvin for pointing out the potential utility of
viewing fermions as anyons, and
M.D.~Kovarik for assistance in programming.
We thank J. Gubernatis and S. Fahy for useful comments,
and David Ceperley for making available a preprint of
Ref.~\cite{OCM}.
Computations were performed at the NCSA under Grant \# DMR-920000N,
and on two IBM RISC workstations in the Solid State Division at
ORNL. This work was supported in part by  the U.S.
Department of Energy through Contract No. DE-AC05-84OR21400
administered by Martin Marietta Energy Systems Inc.
LZ and GSC acknowledge support from the NSF under Grant \# DMR-9101542.

\figure{
Figure 1. Energy $E_0$ vs.~statistics angle $\theta/\pi$,
estimated using the continuous gauge for the statistical `flux'.
The continuous lines give exact eigenvalues, obtained by Lanczos
diagonalization. Sampling variance is indicated by error bars.
We see that, by appropriate choice of weighting of the random walks,
our algorithm can select either the ground state or the lowest excited
state of a different symmetry.

\figure{
Figure 2. Ground-state energy estimate for 6 fermions on a
20x20 lattice, as a function of the time $\tau$, with fixed
sample size ($= 4\times 10^6$ walks).
The horizontal line marks the exact energy.
For clarity, the 1S results are displaced upwards by $+5$.
Some points are marked with parentheses to indicate that the computed
variance is not statistically reliable.
 }

\figure{
Figure 3. Ground-state energy estimate for 4 fermions on a 4x4 lattice,
as a function of $\tau$; sample size is $16\times 10^6$ walks.
Again, the horizontal line gives the exact result.
The zero field case is displaced by $+5$.}

\end{document}